\documentclass[a4paper,10pt,pra,twocolumn]{revtex4}
\usepackage[dvips]{graphics,color}
\usepackage[pdftex]{graphicx}
\usepackage{amsmath}
\usepackage{amsfonts}
\usepackage{amssymb}
\usepackage{amstext}
\usepackage[sort&compress]{natbib}

\begin{document}
\newcommand{\id}{{\sf 1 \hspace{-0.3ex} \rule{0.1ex}{1.52ex}\rule[-.01ex]{0.3ex}{0.1ex}}}
\newcommand{\ignore}[1]{}
%%%%%%%%%%%%%%%%%%%%%%%%%%%%%%%%%%%%%%%%%%%%%%%%%%%%%%%%%
\title{Dephasing assisted transport:
Quantum networks and biomolecules}
%%%%%%%%%%%%%%%%%%%%%%%%%%%%%%%%%%%%%%%%%%%%%%%%%%%%%%%%%
\author{M.~B. Plenio$^{1,2}$ and S.~F. Huelga$^3$}
\affiliation{$^1$ Institute for Mathematical Sciences, Imperial
College London, London SW7 2PG, UK} \affiliation{$^2$ QOLS, Blackett
Laboratory, Imperial College London, London SW7 2BW, UK}
\affiliation{$^3$ Quantum Physics Group, Department of Physics,
Astronomy $\&$ Mathematics\\ University of Hertfordshire, Hatfield,
Herts AL10 9AB, UK}
\date{\today}
\begin{abstract}

Transport phenomena are fundamental in Physics. They allow for
information and energy to be exchanged between individual
constituents of communication systems, networks or even biological
entities. Environmental noise will generally hinder the efficiency
of the transport process. However, and contrary to intuition, there
are situations in classical systems where thermal fluctuations are
actually instrumental in assisting transport phenomena. Here we show
that, even at zero temperature, transport of excitations across
dissipative quantum networks can be enhanced by local dephasing
noise. We explain the underlying physical mechanisms behind this
phenomenon, show that entanglement does not play a supportive role
and propose possible experimental demonstrations in quantum optics.
We argue that Nature may be routinely exploiting this effect and
show that the transport of excitations in light harvesting molecules
does benefit from such noise assisted processes. These results point
towards the possibility for designing optimized structures for
transport, for example in artificial nano-structures, assisted by
noise.

\end{abstract}

\maketitle

{\em Introduction --} Noise is an inevitable feature of any physical
system, be it natural or artificial. Typically, the presence of
noise is associated with the deterioration of performance for
fundamental processes such as information processing and storage,
sensing or transport, in systems ranging from proteins to computing
devices. \\
However, the presence of noise does not always hinder the efficiency
of an information process and biological systems provide a paradigm
of efficient performance assisted by a noisy environment
\cite{neuronas}. A vivid illustration of the counterintuitive role
that noise may play is provided by the phenomenon of stochastic
resonance (SR)\cite{benzi}. Here thermal noise may enhance the
response of the system to a weak coherent signal, optimizing the
response at an intermediate noise level \cite{sr}. Some experimental
evidence suggests that biological systems employ SR-like strategies
to enhance transport and sensing \cite{moss,cell}. Noise in the form
of thermal fluctuations may also lead to directed transport in
ratchets and play a helpful role in Brownian motors
\cite{hanggi,hanggiI,hanggiII}. It seems therefore natural to try
and draw analogies with complex classical networks so that the
physical mechanisms that underpin their functioning when subject to
noise can be perhaps mirrored and eventually used to optimize the
performance of complex quantum networks. Recently, tentative first
steps towards the exploration of the concept of SR in quantum
many-body systems \cite{Viola,us,usI} and quantum communication
channels \cite{Ting,BowenM,DiFranco} have been undertaken
while other studies have focused into analyzing the persistence of
coherence effects in biological systems. In particular, detecting
the presence of quantum entanglement, has been the object of
considerable attention \cite{olaya,Engel}. It was noted, however,
that even if found, it would be unclear whether such entanglement
has any functional importance or is simply the unavoidable
by-product of coherent quantum dynamics in such systems
\cite{BriegelP08}. 

\begin{figure}[t]
\includegraphics[width=0.40\textwidth,height=0.23\textheight]{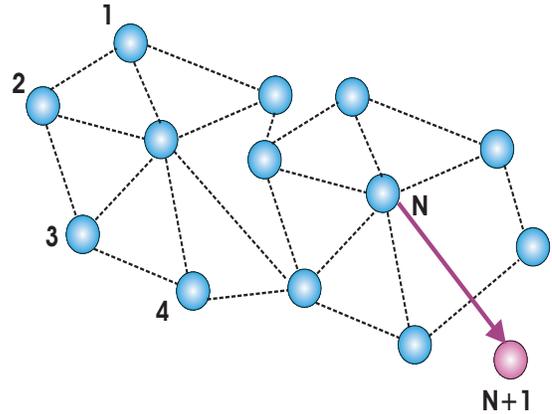}
\caption{Sites (blue spheres), modeled here as spin-1/2 particles or
qubits, are interacting with each other (dashed line) to form a
network. The particles may suffer dissipative losses as well as
dephasing. The red arrow indicates an irreversible transfer of
excitations from the network to a sink that acts as a receiver. }
\label{trans}
\end{figure}
Here we show that dephasing noise, which leads to the destruction of
quantum coherence and entanglement as a result of phase
randomization, may nevertheless be an essential resource to enhance
the transport of excitations when combined with coherent dynamics.
Indeed, we show that a dissipative quantum network subject to
dephasing can exhibit an enhanced capacity for transmission of
classical information when seen as a communication channel, even
though its quantum capacity and quantum coherence are diminished by
the presence of noise. It is the constructive interplay between
dephasing noise and coherent dynamics, rather than the presence of
entanglement, that is responsible for the improved transport
of excitations. Recently, this enhancement of quantum transport 
due to the interplay between coherence and the environment has 
been demonstrated and quantified for chromophoric complexes (see 
\cite{Gaab,lloyd,LloydI,LloydII} and Note Added).\\

In addition to the clarifying nature of these results, it is
intriguing to observe that Nature appears to exploit noise assisted
processes to maximize the system's performance and it will be
worthwhile to explore how similar processes may be useful for the
design of improved transport in nano-structures and perhaps even
quantum information processors.

{\em The basic setting --} We consider a network of $N$ sites that
may support excitations which can be exchanged between lattice sites
by hopping (see Fig. \ref{trans}). The Hamiltonian that describes
this situation is then given by
\begin{equation}
        H = \sum_{k=1}^N \hbar\omega_k \sigma_k^{+}\sigma_k^{-}
        + \sum_{k\neq l} \hbar v_{k,l} (\sigma_k^{-}
        \sigma_{l}^{+} + \sigma_k^{+}\sigma_{l}^{-}),
        \label{NetworkHamiltonian}
\end{equation}
where $\sigma_k^{+}$ ($\sigma_k^{-}$) are the raising and lowering
operators for site $k$, $\hbar\omega_k$ is the local site excitation
energies and $v_{k,l}$ denotes the hopping rate of an excitation
between the sites $k$ and $l$. It should be noted that the dynamics
in this system preserves the total excitation number in the system.
%operator $\sum_{k=1}^N \sigma_k^{+}\sigma_k^{-}$.
This is not an essential
feature but makes the system amenable to efficient numerical
analysis. We will assume that the system is susceptible
simultaneously to two distinct types of noise processes, a
dissipative process that reduces the number of excitations in the
system at rate $\Gamma_k$ and a dephasing process that randomizes
the phase of local excitations at rate $\gamma_k$.

Initially we will assume that we can describe both processes by
using a Markovian master equation with local dephasing and
dissipation terms. It is important to note however that the effects
found here persist when taking
account of the system-environment interaction in a more detailed
manner (see Methods). Dissipative processes, which lead
to energy loss, are then described by the Lindblad super-operator
\begin{equation}
        {\cal L}_{diss}(\rho) = \sum_{k=1}^{N} \Gamma_k[
        -\{\sigma_k^{+}\sigma_k^{-},\rho\} +
        2 \sigma_k^{-}\rho \sigma_k^{+} ],
\end{equation}
while energy-conserving dephasing processes are described by the
operator
\begin{equation}
        {\cal L}_{deph}(\rho) = \sum_{k=1}^{N} \gamma_k[
        -\{\sigma_k^{+}\sigma_k^{(-)},\rho\} +
        2 \sigma_k^{+}\sigma_k^{-}\rho \sigma_k^{+}\sigma_k^{-}].
\end{equation}
Finally, in order to be able to measure the total transfer of
excitation, we designate an additional site, numbered $N+1$, which
is populated by an irreversible decay process from a chosen level
$k$ as described by the Lindblad operator
\begin{eqnarray}
        {\cal L}_{sink}(\rho) &=& \label{Lindblad}\\
        && \hspace*{-1.5cm}\Gamma_{N+1}[
        -\{\sigma_k^{+}\sigma_{N+1}^{-}\sigma_{N+1}^{+}
        \sigma_k^{-},\rho\} +
        2 \sigma_{N+1}^{+}\sigma_k^{-}\rho \sigma_k^{+}\sigma_{N+1}^{-} ].
        \nonumber
\end{eqnarray}
The subindex {\rm 'sink'} emphasizes that no population can escape of
site $N+1$. For definitiveness and simplicity, the initial
state of the network at $t=0$ will be assumed to be a single
excitation in site $1$ unless stated otherwise.

The key question that we will pose and answer is the following: {\em
In a given time $T$, how much of the initial population in site $1$
will have been transferred to the sink at site $N+1$ and how is this
transfer affected by the presence of dephasing and dissipative
noise.}

In the remainder of this paper we will demonstrate that, in certain
settings, the presence of dephasing noise can assist the transfer of
population from site $1$ to the sink at site $N+1$ considerably. It
is an intriguing observation that this noise enhanced transfer does
not occur for all possible Hamiltonians of the type given by
eq.(\ref{NetworkHamiltonian}) and may depend also on properties of
the noise such as its spatial dependence. These noise rates can be
optimized numerically, and in very simple cases analytically, to
yield the strongest possible effect. One may suspect that
natural, biological systems, have actually made use of such an
optimization.

{\em Linear chain --} We begin with a brief analysis of the uniform
linear chain with only nearest neighbor interactions so that in eq.
(\ref{NetworkHamiltonian}) the coupling strengths satisfy $v_{l,k} =
v_{k,l} = v \delta_{l,k+1}$ for $k=1,\ldots,N-1$ and $\omega_k =
\omega$ and $\Gamma_k = \Gamma$ for $k=1,\ldots,N$. Extensive
numerical searches show that, for arbitrary choices of
$\Gamma_{N+1}$, $\Gamma$ and $\omega$ and arbitrary transmission
times $T$ and chains of the length $N = 2,\ldots,12$, the optimal
choice of dephasing noise rates vanish. We have used a directed
random walk algorithm with multiple initial states which has never
exceeded the values for the noise-free chain and approached them to
within at least $10^{-8}$.
%
%Obtaining a general proof of this observation will certainly be
%difficult but
We were able to derive formulae for the case
$T=\infty$ and short chains which demonstrate this behaviour
analytically. For $N=2$, with $\omega_1=\omega_2=\omega$ and
arbitrary $v_{1,2}$, $\gamma_i$ and $\Gamma_i$, we find, with the
abbreviation $\gamma=\gamma_1+\gamma_2$ and $x =
2\Gamma_1^3+\Gamma_1\Gamma_3(3\Gamma_1+\Gamma_3)$, that the
population of the sink is given by
\begin{equation}
        p_{sink} = \frac{\Gamma_3 v_{1,2}^2}{x+ \Gamma_1(\Gamma_1+\Gamma_3)\gamma
        +(\Gamma_3+2\Gamma_1)v_{1,2}^2 },
\end{equation}
which is evidently maximized for $\gamma=0$. One may also obtain the
analytical expressions for $N=3$ and $\Gamma_k=\Gamma$ for $k=1,2,3$
and demonstrate that the optimal dephasing level is $\gamma=0$ (see
section on {\em Methods}). This approach, though more tedious, may
be taken to higher values of $N$ as well. Extensive numerical
searches lend further support to the observation that dephasing does
not improve excitation transfer for uniform chains but a general
proof has remained elusive.

So far, the findings are consistent with the expectation that noise
does not enhance the transport of excitations. However, for
non-uniform chains we encounter the different and perhaps surprising
situation where noise can significantly enhance the transfer rate of
excitations.

\begin{figure}[t]
\includegraphics[width=0.47\textwidth,height=0.23\textheight]{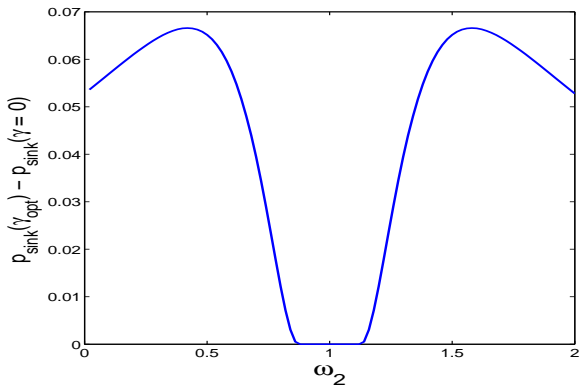}
\caption{The optimal improvement of the transfer efficiency is
plotted versus the site frequency $\omega_2$ in a chain of length
$N=3$ and system parameters $\omega_1=\omega_3=1$,
$\Gamma_1=\Gamma_2=\Gamma_3=1/100$, $v_{1,2}=v_{2,3}=1/10$,
$\Gamma_{N+1}=1/5$ and $T=\infty$. One observes that dephasing only
assists the transmission probability in some frequency intervals.}
\label{Improvement2}
\end{figure}

\begin{figure}[t]
\includegraphics[width=0.45\textwidth,height=0.23\textheight]{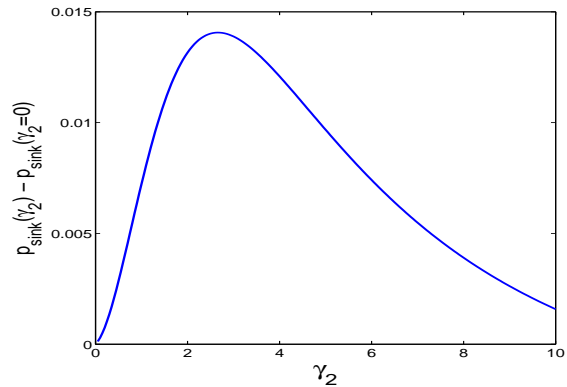}
\caption{The difference between transfer efficiency and the
efficiency without dephasing is plotted versus the dephasing rate
$\gamma_2$ in a chain of length $N=3$ and
$\omega_1=\omega_2/4=\omega_3=1$, $v_{1,2}=v_{2,3}=1/10$,
$\gamma_1=\gamma_3=0$, $\Gamma_k = 1/100$ for $k=1,\ldots, N$,
$\Gamma_{N+1}=1/5$ and $T=\infty$. Initially increasing dephasing
assists the transfer of excitation while very strong dephasing
suppresses the transport.} \label{Improvement3}
\end{figure}

As an illustrative example, we may keep the nearest neighbor
coupling uniform but allow for one site to have a different site
energy $\omega$. If we chose $N=3$, $\omega_1=\omega_3=1$,
$\Gamma_1=\Gamma_2=\Gamma_3=1/100$, $v_{1,2}=v_{2,3}=1/10$,
$\Gamma_{N+1}=1/5$ and $T=\infty$ , then we obtain the results
depicted in Fig. \ref{Improvement2}. One observes that dephasing
assists the transmission only when site $2$ is sufficiently detuned
from the neighboring sites.
This example suggests a simple picture to explain the reason for the
dephasing enhanced population transfer through the chain. Site $2$
is strongly detuned from its neighboring sites and the coupling $v$
to its neighbors is comparatively weak, i.e. $v\ll \delta\omega$
with $\delta\omega =\min[|\omega_2-\omega_1|,|\omega_3-\omega_2|]$.
Hence , the transport rate is limited by a quantity of order
$v^2/\delta\omega$ as it is a second order process due to the lack
of resonant modes between neighboring sites. Introducing dephasing
noise leads to a broadening of the energy level at each site $k$ and
a line-width proportional to the dephasing rate $\gamma_k$. Then,
with increasing dephasing rate, the broadened lines of neighboring
sites begin to overlap and the population transfer will be enhanced
as resonant modes are now available. Enhancing the dephasing rate
further will eventually lead to a weakening of the transfer as the
modes are distributed over a very large interval and resonant modes
have a small weight. Dissipation does not lead to the same
enhancement as, crucially, the gain to the broadening of the line is
overcompensated by the irreversible loss of excitation. This is
corroborated by numerical studies where increasing dissipation does
not assist the transport.
The physical picture outlined above is confirmed in Fig.
\ref{Improvement3}. We chose a chain of length $3$ which suffers
dephasing only in site $2$ and uniform dissipation with rates
$\Gamma_k = 1/100$ along the chain while
$\omega_1=\omega_2/4=\omega_3=1$ and $v_{1,2}=v_{2,3}=1/10$ (see
fig. \ref{Improvement3}). The close relationship of this model to
Raman transitions in quantum optics will be exploited to propose a
realizable experiment in a highly controlled environment to verify
these effects (see section on realizations).

In the examples above the improvement of excitation transfer due to
the dephasing is small. One can easily show, however, that this
improvement may be made arbitrarily large in the sense that without
noise the transfer rate approaches zero while it approaches unity
arbitrarily closely for optimal noise levels. As an example, for
$N=3,\omega_1=\omega_3=1;\omega_2=100,v_{1,2}= v_{2,3}=v$,
$\gamma_1=\gamma_3=0$ and $\Gamma_1=\Gamma_2=\Gamma_3=v^2/f$ and
$\Gamma_4 = 10^5v$ we find for $\Delta
p=p_{sink}(\gamma_{2,opt})-p_{sink}(\gamma_2=0)$ that
\begin{equation}
        \lim_{v\rightarrow 0} \Delta p =
        \frac{f^2 \gamma_2^2}{f^2\gamma_2^2 + 
        3f\gamma_2((\omega_2-1)^2+\gamma_2^2) + 
        ((\omega_2-1)^2+\gamma_2^2)^2}
\end{equation}
This is maximized for $\gamma_2=\omega_2-1$ when it takes the value
$\Delta p = f^2/(f^2+6f(\omega_2-1)+4(\omega-2-1)^2)$. In the limit
$f\rightarrow\infty$ this approaches $1$, that is, without noise the
excitation transfer vanishes while with noise it achieves unit
efficiency! It should be noted that being a system of fixed finite
size, the effect may not be directly attributed to
Anderson localization \cite{Anderson} which, in addition does not
occur in systems attached to a sink, as is assumed here
\cite{Gurvitz00}.

{\em Entanglement and coherence in the channel --} We have
seen that the transport of excitations in the system may be
assisted considerably by local dephasing. Now we would like
to discuss briefly the quantum coherence properties during
transmission by studying the presence of entanglement and
the ability of the chain to transmit quantum information.
To this end, we consider how entanglement is transported along the
chain when it is used to propagate one half of a maximally entangled
state to obtain an insight on how is the quantum capacity of this
channel affected by dephasing. To illustrate this, we consider a
chain of $N=4$ sites. We chose the same parameters as in Fig.
\ref{Improvement2} and fix $\omega_3=14$. Comparison of the
entanglement between an uncoupled site and the various sites in the
chain for vanishing dephasing and the optimal choice of the
dephasing for excitation transfer show that, while entanglement
propagates through the system, the amount of entanglement decreases
with increasing dephasing. In fact, the dephasing rate that
optimizes the ability of the channel to transmit quantum information
vanishes, in contrast to the situation for excitation transfer.
Therefore, although dephasing may enhance the propagation of
excitations, it also destroys quantum coherence and in the present
setting it leaves an overall detrimental effect.

{\em Complex networks and Light-harvesting molecules --} So far, we have
demonstrated that in linear chains local dephasing noise may enhance
the transfer of excitations. Going beyond this, we will now consider
fully connected networks and apply our observations to a
model that describes the transfer of excitons in
the Fenna-Matthews-Olson complex of {\em Prosthecochloris
aestuarii}, which is a pigment-protein complex that consists of
seven bacteriochlorophyll-a (BChla) molecules (see
\cite{lloyd,LloydI,LloydII} and Note Added for closely related 
work). This complex is able
to absorb light to create an exciton. This exciton then propagates
through the complex until it reaches the {\em reaction centre} where
its energy is then used to trigger further processes that bind the
energy in chemical form \cite{WikiFMO,AdolphsR06}. The Hamiltonian
of this complex may be approximated by eq.
(\ref{NetworkHamiltonian}), where the site energies and coupling
constants may be taken from table 2 and 4 of \cite{AdolphsR06}. We
then find, in matrix form
\begin{equation}
        H \!=\!\! \left(\!\!\begin{array}{rrrrrrr}
         215   & \!-104.1 & 5.1  & -4.3  &   4.7 & -15.1 &  -7.8 \\
        \!-104.1 &  220.0 & 32.6 & 7.1   &   5.4 &   8.3 &   0.8 \\
           5.1 &   32.6 &  0.0 & -46.8 &   1.0 &  -8.1 &   5.1 \\
          -4.3 &    7.1 &\!-46.8 & 125.0 &\! -70.7 &\! -14.7 &  -61.5\\
           4.7 &    5.4 &  1.0 & \!-70.7 & 450.0 &  89.7 &  -2.5 \\
         -15.1 &    8.3 & -8.1 & -14.7 &  89.7 & 330.0 &  32.7 \\
          -7.8 &    0.8 &  5.1 & -61.5 &  -2.5 &  32.7 & 280.0
          \end{array}\!\!
        \right)
\end{equation}
where we have shifted the zero of energy by $12230$ (all number are
given in the units of $1.988865\cdot 10^{-23}Nm =
1.2414\,10^{-4}eV$) for all sites corresponding to a wavelength of
$\cong 800nm$. Recent work \cite{AdolphsR06} suggests that it is
this site $3$ that couples to the reaction centre at site 8. For
this rate, somewhat arbitrarily, we chose $\Gamma_{3,8} = 10/1.88$
corresponding to about 1 $ps^{-1}$ (value in the literature
range from 0.25$ps^{-1}$ \cite{AdolphsR06} and 1 $ps^{-1}$ \cite{LloydII} 
to 4 $ps^{-1}$ \cite{olaya}). 
Again, we will assume the
presence of both dissipative noise (loss of excitons) and dephasing
noise (due to the presence of a phonon bath consisting of
vibrational modes of the molecule). The measured lifetime of
excitons is of the order of 1 $ns$ which determines a dissipative
decay rate of $2\Gamma_k = 1/188$ and that we assume to be the same
for each site \cite{AdolphsR06}. If we neglect the presence of any
form of dephasing and we start with a single excitation on site $1$,
then we observe that the excitation is transferred to the reaction
centre (site 8). For a time $T=5$, we find that the amount of
excitation that is transferred is $p_{sink}=0.551926$. Optimal
dephasing rates that maximize the transfer rate of the initial
excitation in site $1$ considerably improve on that. For $T=5$ we
find the optimal dephasing rates
$(\gamma_1,\gamma_2,\gamma_3,\gamma_4,\gamma_5,\gamma_6, \gamma_7)=
(469.34, 5.36, 99.13, 5.55, 114.86, 1.88, 291.08)$ and the much
improved value $p_{sink}=0.988526$. For $T=\infty$, we find the
dephasing free transfer probability of $p_{sink}=0.81425$ while for
the optimal dephasing rates $(\gamma_1,\gamma_2,
\gamma_3,\gamma_4,\gamma_5,\gamma_6,\gamma_7)=(27.40,26.84, 1.22,
87.12, 99.59, 232.76,88.35)$ we find $p_{sink}=0.99911$. It should
be noted that these dephasing rates are comparable to the inter-site
coupling rates which suggests that a more accurate treatment will
need to go beyond master equations (see Methods for a brief discussion).
\begin{figure}[t]
\includegraphics[width=0.47\textwidth,height=0.23\textheight]{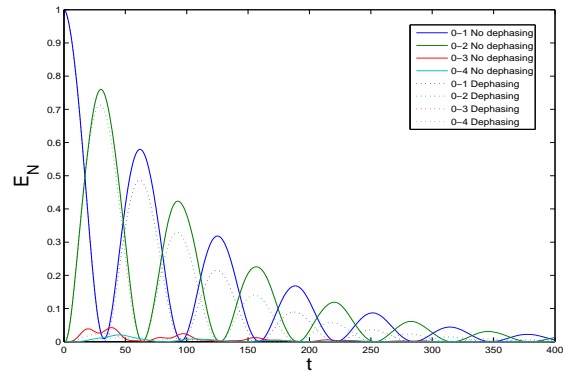}
\caption{The time evolution of the entanglement between a decoupled
site and the sites in the chain of length $N=4$ and system
parameters $\omega_1= \omega_2=\omega_4=10$, $\omega=14$,
$v_{1,2}=v_{2,3}=v_{3,4}=1$, $\Gamma_k = 1/10$ for $k=1,\ldots, N$
and $\Gamma_{N+1}=1$. The initial state is a maximally entangled
state between the decoupled site and the first site of the chain.
Dephasing destroys entanglement along the chain and has no
beneficial effect. } \label{Improvement}
\end{figure}

We conclude that dephasing may lead to a very strong enhancement of
the transfer rate of excitations in a realistic network. In fact, in
models obtained from spectroscopic data measured on the FMO complex
it is indeed observed that almost complete transport should take
place within time $T=5$ \cite{AdolphsR06}. It is remarkable that
such a rapid transfer cannot be explained from a purely coherent
dynamics and, as shown above, the underlying reason for the speed up
is the presence of dephasing which may even be local.

\begin{figure}[t]
\includegraphics[width=0.40\textwidth,height=0.20\textheight]{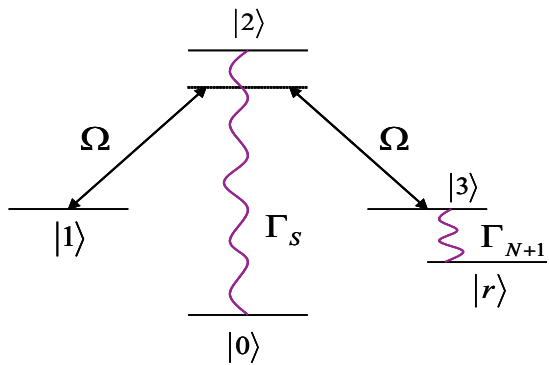}
\caption{A atomic system with Raman transitions provides a
transparent illustration of dephasing assisted transport. The
required level structure may be realized in $Ca^{+}, Sr^{+}$ or
$Ba^{+}$. Each atomic level represents a site in the chain which may
be populated. Starting with all the population in level $1$, one may
then irradiate the system with classical laser fields of
Rabi-frequency $\Omega$ on the $1\leftrightarrow 2$ and the
$3\leftrightarrow 2$ transition \cite{ScullyZubairy}. Level 3 in
turn is assumed to decay spontaneously into an additional level
$|r\rangle$ that plays the role of the recipient. Spontaneous decay
of the chain as a whole is modelled by spontaneous decay into level
$|0\rangle$ from which no population can enter the levels
$|1\rangle,|2\rangle,|3\rangle$ and $|r\rangle$ anymore. Dephasing
noise may now enter the system affecting level $2$ for example
through magnetic field fluctuations.} \label{Realization}
\end{figure}

{\em Experimental Realizations --} The FMO-complex provides a
fascinating setting for the observation of dephasing enhanced
transport but it is also a very challenging environment to
verify the effect precisely. Here we present several physical
systems in which the dephasing enhanced excitation transfer
may be observed and which are at the same time highly controllable.
Perhaps the simplest such
setting is found in atomic physics (see Fig. \ref{Realization})
where the behaviour of a chain of three sites may be simulated using
detuned Raman transitions in ions such as $Ca^{+}, Sr^{+}$ or
$Ba^{+}$.
% Each atomic level represents a site in the chain
% which may be populated. Starting with all the population in level
% $1$, one may then irradiate the system with classical laser fields
% of Rabi-frequency $\Omega$ on the $1\leftrightarrow 2$ and the
% $3\leftrightarrow 2$ transition \cite{ScullyZubairy}. Level 3 in
% turn is assumed to decay spontaneously into an additional level
% $r\rangle$ that plays the role of the recipient. Spontaneous decay
% of the chain as a whole is modelled by spontaneous decay into level
% $|0\rangle$ from which no population can enter the levels
% $|1\rangle,|2\rangle,|3\rangle$ and $|r\rangle$ anymore.
% Dephasing noise may now enter the system affecting level $2$ for
% example through magnetic field fluctuations.
The master equation of this system simulates exactly that of a chain
with a single excitation as has been described throughout this
paper. Atomic populations may be measured with very high accuracy
using quantum jump detection \cite{PlenioK98,QJump}.

A variety of other natural implementations of dephasing
assisted excitation transport can be conceived and will
be studied in detail elsewhere. Firstly,
the oscillations of ions in a linear ion trap transversal
to the trap axis realizes a harmonic chain \cite{PlenioHE04}
that allows for the implementation of a variety of operations
such as preparation of Fock states and is capable of supporting
nearest neighbor coupling between neighbouring ion oscillators
\cite{SerafiniRP07} and allowing high efficiency readout by
quantum jump detection \cite{PlenioK98}. When restricting to
the single excitation space, the dynamics of the system is
described by master equations that become equivalent to those
presented in this paper.

Furthermore, harmonic chains are also realized in coupled arrays of
cavities which have recently received considerable attention in the
context of quantum simulators \cite{HartmannBP06}. Ultra-cold atoms
in optical lattices which have previously been used to study
thermal assisted transport in Brownian ratchets \cite{LatticeRatchet}
presents
another scenario in which to study such dephasing assisted processes.
Chains of superconducting qubits or superconducting stripline cavities
\cite{scq} may also provide a possible setting for the observation
of the effects described above.

{\em Conclusions and outlook --}
% From the viewpoint of quantum information theory, strongly
% interconnected networks such as those discussed above may
% also be considered as a quantum channel for the transmission
% of classical or quantum information. We have shown that dephasing
% noise can dramatically increase the classical capacity of
% such a quantum channel, while quantum correlations do degrade when
% the system is subject to dephasing noise.
The results presented here demonstrate that while dephasing noise
destroys quantum correlations, it may at the same time enhance the
transport of excitations. In fact, the efficient transport observed
in certain biological systems has been shown to be incompatible with
a fully coherent evolution while it can be explained if the system
is subject to local dephasing. Hence, in this context, the presence
of quantum coherence and therefore, entanglement in the system, does
not seem to be supporting excitation transfer. This suggests that
entanglement that may be present in bio-molecules, though
interesting, may not be a universal functional resource.

Importantly, the results presented here suggest that it may be
possible to design and optimize the performance of nano-fabricated
transmission lines in naturally noisy environments to achieve
strongly enhanced transfer efficiencies employing the concept
of noise assisted transport.

{\em Acknowledgements--} We are grateful to Seth Lloyd for 
helpful communications concerning \cite{lloyd,LloydI,LloydII},
Neil Oxtoby, Angel Rivas and Shashank Virmani for useful comments on 
the manuscript and to Danny Segal for advice on atomic physics. 
This work was supported by the EU via the Integrated
Project QAP ({\rm `Qubit Applications'}) and the STREP action CORNER
and the EPSRC through the QIP-IRC. MBP holds a Wolfson
Research Merit Award.\\

{\em Note Added---} While finalizing this work, we became aware of
independently obtained but closely related results presented in 
\cite{lloyd,LloydI,LloydII}.
There it was showed that quantum transport can be enhanced by an 
interplay between coherent dynamics and environment effects with 
particular emphasis on excitonic energy transfer in light harvesting 
complexes \cite{LloydII}. The role of the different physical processes 
that contribute to the energy transfer efficiency have been studied 
in \cite{LloydI} and the enhancement of quantum transport due to a 
pure dephasing environment within the Haaken-Strobl model
was demonstrated in \cite{lloyd}.

\newpage

{\em Methods --}\\

{\em Exact solutions for uniform chains --}
One may also obtain the analytical expressions
for a chain of length $N=3$ described by eqs. (\ref{NetworkHamiltonian})
- (\ref{Lindblad}) for the choice and $\Gamma_k=\Gamma$ for $k=1,2,3,4$
and demonstrate that the optimal dephasing level is $\gamma=0$.
We find
\begin{widetext}
\begin{displaymath}
        p_{sink} = \frac{(4\Gamma+\gamma_1+\gamma_3)v^2}{36\Gamma^5
        + 6a\Gamma^4+2\Gamma^3(3\gamma_1^2+3\gamma_2^2+8b+2\gamma_3^2
        +32v^2)+ \Gamma^2(2c +d v^2)
        +\Gamma v^2(3\gamma_1^2+7b+4\gamma_3^2+15v^2)
        +4(\gamma_1+\gamma_3)v^4}
\end{displaymath}
\end{widetext}
where $a=(5\gamma_1+5\gamma_2+4\gamma_3)$, $b = \gamma_1\gamma_2+
\gamma_1\gamma_3+\gamma_2\gamma_3$,
$c=\gamma_1(\gamma_2^2+\gamma_3^2)
+\gamma_2(\gamma_1^2+\gamma_3^2)+\gamma_3(\gamma_1^2+\gamma_2^2)
+2\gamma_1\gamma_2\gamma_3$, $d=32\gamma_3+25\gamma_2+29\gamma_1$.
Then one first observes that the optimal choice is $\gamma_2=0$ as
it only occurs in the denominator with positive coefficients. In the
remaining expression one then substitutes
$\gamma_k={\tilde\gamma}_k^2$ allowing also for negative
${\tilde\gamma}_k$. Then differentiation w.r.t these
${\tilde\gamma}_k$ shows that the gradient only vanishes for
${\tilde\gamma}_1={\tilde\gamma}_2=0$.

{\em Beyond Markovian master equations--} So far we have
demonstrated the existence of dephasing enhanced excitation
transfer employing a master equation description. The optimized
dephasing rates that have been obtained, in particular
those in the context of the FMO complex, can be comparable
to the coherent interaction strengths and may be similar
to the spectral width of the bath responsible for the
dephasing \cite{AdolphsR06}. This may not be fully compatible
with the master equation approach employed so far as its
derivation relies on several assumptions including the weak
coupling hypothesis and the requirement for the bath to be
Markovian \cite{BreuerP02}. The derivation is further complicated
for systems with several constituents where the local coupling
of its constituents is not compatible with non-local structure of
the eigenmodes of the systems. This is especially so when the
coherent inter sub-system coupling is of comparable strength
to the system environment coupling. The situation is made more
difficult due to spatial as well as temporal correlations in the
environmental noise (which is to be expected in particular for the
FMO complex but also many other realisations of coupled chains in
contact with an environment). Bloch-Redfield equations and other
effective description are sometimes used but still represent
approximations to the correct dynamics \cite{BreuerP02} where the
errors are often difficult to estimate precisely.
\begin{figure}[b]
\includegraphics[width=0.47\textwidth,height=0.23\textheight]{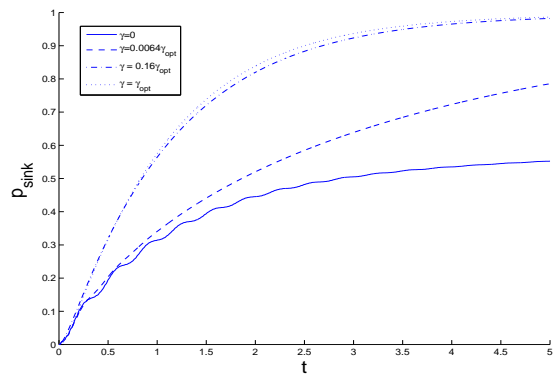}
\caption{Here we show how the transfer in the presence of dephasing
into a bath that is modelled by a collisional model where local
sites briefly interact with a single particle. The interaction
strength is chosen such that in an uncoupled systems the sites
suffer the optimal decoherence rates $\gamma_{opt}$ as presented in
the previous section multiplied with factors $0,0.0064,0.16$ and
$1$. The dynamics is similar to that observed for the master
equation approach and shows only minor deviations. Increased
dephasing rates do improve the excitation transfer also in this
model.} \label{Real bath}
\end{figure}
% Apart from the
% possibility of spatial as well as temporal correlations in the
% environmental noise (which is to be expected in particular for
% the FMO complex but also many other realisations of coupled chains
% in contact with an environment), the existence of interactions
% between constituents leads to the question whether the Lindblad
% operators will act in the local basis or globally on eigenmodes
% of the system. In fact, which description is correct will depend
% on a variety of properties of system and environment and their
% interaction. This is especially so when the interaction between
% system and environment gives rise to noise rates that are of the
% order of the coupling strength between subsystems.

Therefore, we demonstrate briefly that dephasing assisted transfer
of excitation can also be observed when one uses a microscopic model
of an environment that may, in addition, exhibit non-Markovian
behaviour. To this end we study the effect of an environment which
is modelled by brief interactions between two-level systems and
individual subsystem of the chain in which excitation transport is
taking place. The strength and nature of the interactions can be
chosen to implement dephasing (elastic collisions) and dissipation
(in-elastic collisions). Non-markovian effects can be included in
the model depending on the spatial and temporal memory of the
environment particles. Interaction strengths are determined for a
single site system to obtain the dissipation rate $\Gamma$ and
dephasing rate $\gamma$. This simplified model allows us to study
the effect of more realistic environments outside the master
equation picture and results are summarized in Figure 6. A more
detailed simulation of excitation transfer taking account of the
full environment are beyond the scope of the present work and will
be presented elsewhere \cite{Environment}

\end{document}